\documentclass[aps,pre,numerical,reprint,amsmath,amssymb]{revtex4-1}

\usepackage{graphicx}
\usepackage{dcolumn}
\usepackage{bm}
\usepackage{color}
\usepackage{soul}

\begin{document}

\newcommand{\Froude}{\mbox{\textit{Fr}}} 

\graphicspath{{figures/}}

\title{Volume entrained in the wake of a disc intruding into an oil-water interface}

\author{Ivo R. Peters$^{1,2}$}
\email{i.r.peters@soton.ac.uk}
\author{Matteo Madonia$^1$}
\author{Detlef Lohse$^1$}
\author{Devaraj van der Meer$^1$}
\affiliation{
$^1$Physics of Fluids, Faculty of Science and Technology and J.M. Burgers Centre for Fluid Dynamics, \\
P.O. Box 217, 7500 AE Enschede, The Netherlands\\
$^2$Engineering and the Environment, University of Southampton, Highfield, Southampton SO17 1BJ, UK
}

\date{\today}

\begin{abstract}
An object moving through a plane interface into a fluid deforms the interface in such a way that fluid from one side of the interface is entrained into the other side, a phenomenon known as Darwin's drift. We investigate this phenomenon experimentally using a disc which is started exactly at the interface of two immiscible fluids, namely oil and water. First, we observe that due to the density difference between the two fluids the deformation of the interface is influenced by gravity, and show that there exits a time window of universal behavior. Secondly, we show by comparing with boundary integral simulations that, even though the deformation is universal, our results cannot be fully explained by potential flow solutions. We attribute this difference to the starting vortex, which is created in the wake of the disc. Universal behavior is preserved, however, because the size and strength of the vortex shows the same universality as the potential flow solution.
\end{abstract}

\pacs{}

\maketitle

\section{Introduction}

The flow around objects in a bulk medium is a classical problem in fluid dynamics which occurs in numerous natural phenomena and has many practical applications. The majority of research concerns high Reynolds number flows, where vortex shedding plays an important role, and mixing is greatly enhanced~\cite{Choi2008}. Until recently, and even though it was pointed out more than 60 years ago, much less attention has been given to mixing due to the drift volume, a mechanism described by potential flow known as Darwin's drift~\cite{Darwin1953}. One of the recent studies has shown that for many marine animals Darwin's drift is the dominant contribution to mixing by swimming~\cite{Katija2009}. Darwin's original proposition states that the drift volume induced by an object, which was started infinitely far away from an imaginary plane and moves infinitely far to the other side of the plane is equal to the added mass volume of that object~\cite{Darwin1953}. This prediction is based on potential flow, and has been confirmed two decades ago by Eames \emph{et al.}~\cite{Eames1994}, who added specific information concerning the method of evaluating the integrals to calculate the drift volume and of taking into account partial drift for finite systems. An open question is how the drift volume is influenced in the case of a less smooth object, or at higher Reynolds numbers, when flow separation starts to play a role. This problem is related to the drift volume of a vortex ring, which was considered analytically for a continuously expanding vortex~\cite{Turner1964} and experimentally for a steady vortex ring~\cite{Dabiri2006}.

We approach this problem experimentally by impulsively starting a disc from the interface of two immiscible liquids, oil and water, similarly and with the same setup as we did before when studying the dynamics of an air-water interface when a disc would impact it~\cite{Bergmann2006,Bergmann2009,Gekle2009a,Gekle2010,Gekle2010d}. Besides being a practical solution to studying drift volumes, this method has relevance for, e.g., the mechanism of oil dispersion in water~\cite{Murphy2015}. In our experiment, when the disc moves down, it drags along the oil, which then obtains a particular funnel-shaped profile (see, e.g., Fig.~\ref{fig:sequence_thick_layer}). Both gravity and surface tension are deforming the shapes, and we determine how these profiles depend on the velocity of the disc. We observe that there exists a universal profile which becomes more prominent for higher velocities, for which this universal shape will extend deeper into the fluid. Surprisingly, however, despite the observed universal behavior, these profiles do not agree with potential flow simulations we performed using a boundary integral technique. We attribute this difference to the formation of a vortex ring, which distorts the potential flow. We specifically show that the drift volume is \emph{larger} than that predicted by potential flow.

\section{Experimental setup}
\begin{figure}
    \centering
    \includegraphics[width=70mm]{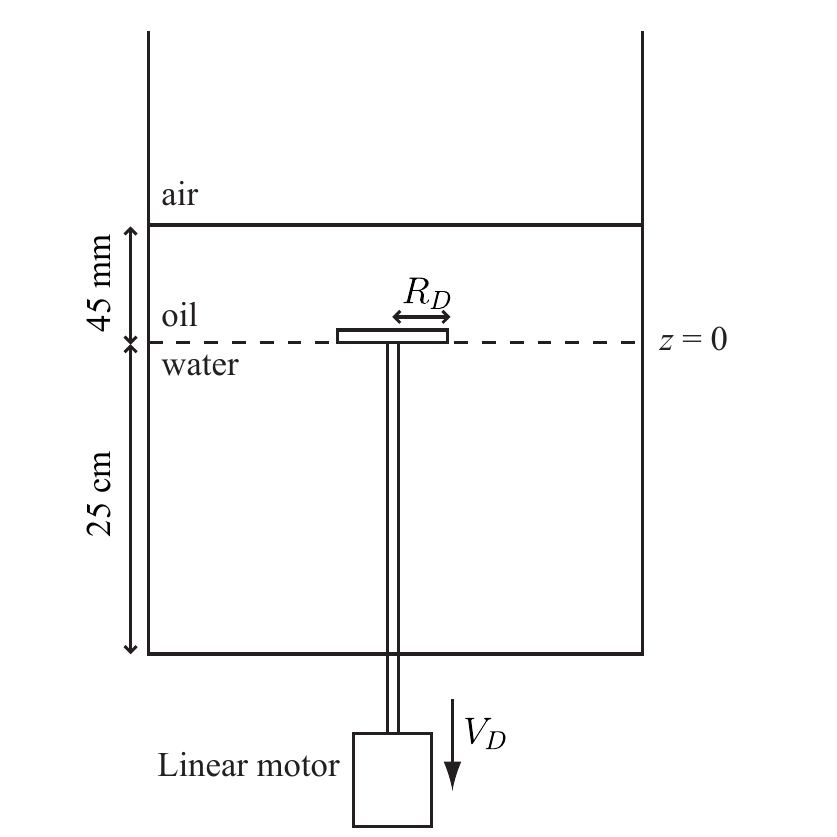}
    \caption{\label{fig:setup} Schematic view of the experiment, with disc radius $R_D$ and disc velocity $V_D$. We start with the bottom of the disc at rest at the interface between a deep layer of oil (45 mm) on top of a deep layer of water (25 cm), after which we pull down the disc at constant speed $V_D$. We define $z=0$ at the undisturbed oil-water interface.}
\end{figure}
The experimental setup (Fig.~\ref{fig:setup}) consisted of a water reservoir with a cross section of 15 cm by 15 cm and a height of 50 cm. A linear motor that was mounted below the tank pulls a disc with a radius $R_D=20~\mathrm{mm}$ through the water surface at a constant speed $V_D$, by means of a thin rod (radius 3 mm) connecting the linear motor with the disc. The disc was accelerated with a maximum acceleration of $42~\mathrm{m/s^2}$ until the desired velocity ($V_D$) was reached. The events were recorded with a Photron SA2 high-speed color camera at frame rates ranging from $1$ to $8~\text{kHz}$. The main control parameter of the experiment is an effective Froude number $\Froude^*$, which is similar to the regular Froude number defined as the disc speed $V_D$, made dimensionless using the disc radius $R_D$ and the gravitational acceleration $g$. Only we replace $g$ by the effective gravitational acceleration $g^*$ of the oil phase inside the water phase, as one would use to determine the wave speed of gravitational waves on a density interface $g^*=g(\rho_w-\rho_o)/(\rho_w+\rho_o)$ \cite{Kundu2004_waves}, yielding
\begin{equation}
    \Froude^*=\frac{V_D^2}{gR_D}\left(\frac{\rho_w+\rho_o}{\rho_w-\rho_o}\right),
\end{equation}
where $\rho_w$ and $\rho_o$ are the densities of water and oil respectively. In our experiments we used sunflower oil, which has a density $\rho_o=900~\mathrm{kg/m^3}$ and a viscosity $\nu\sim50\cdot10^{-6}\mathrm{m^2/s}$. Next to demineralized water we used a solution of table salt in water to increase the density of the water phase. We dissolved $1.0~\mathrm{kg}$ of table salt in $5000~\mathrm{ml}$ water, resulting in $\rho_{sw}=1140~\mathrm{kg/m^3}$. The thickness of the oil layer in these experiments was $45~\mathrm{mm}$, which was thick enough to be considered as infinite. We verified this by performing the same experiment with increased oil layer thicknesses of $90~\mathrm{mm}$ and $135~\mathrm{mm}$, which did not influence our results.

\section{Results}
\begin{figure}[htb]
    \centering
    \includegraphics[width=86mm]{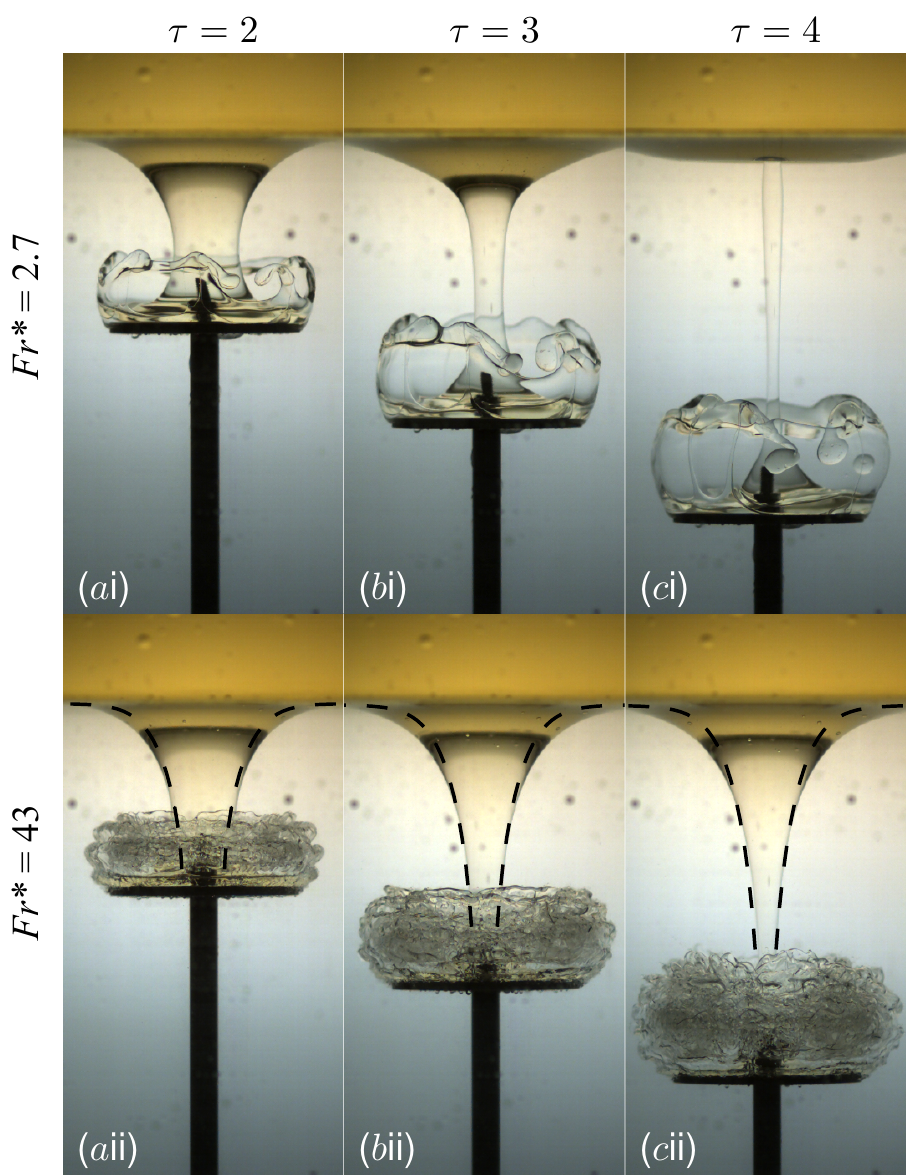}
    \caption{\label{fig:sequence_thick_layer}
    Snapshots from experiments with two different values of the effective Froude number $\Froude^*=2.7$ ($a$i-$c$i), and $\Froude^*=43$ ($a$ii-$c$ii). Corresponding pictures have been taken at the same dimensionless times $\tau=2,~\tau=3,~\tau=4$. In the top experiment gravity has a clear influence on the shape of the entrained oil column. The bottom experiment is in the inertial regime, where gravity has negligible influence. A vortex ring is formed above the disc, which grows with the same dimensionless rate, independent of the Froude number. Black dashed line is the result from boundary integral simulations.
    }
\end{figure}
\begin{figure}
    \centering
    \includegraphics[width=85mm]{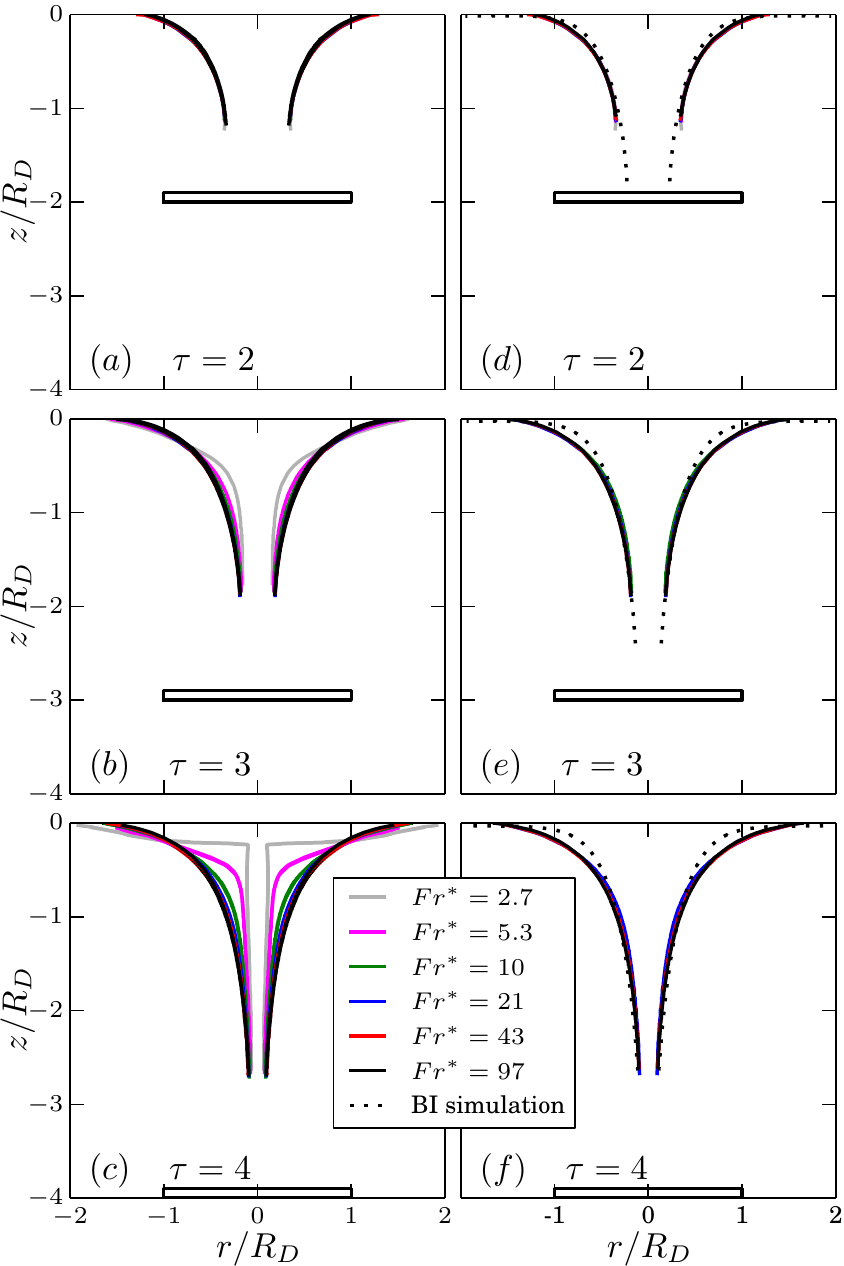}
    \caption{\label{fig:all_profiles}
    Profiles of the entrained oil, for effective Froude numbers ranging from $2.7$ to $97$. We observe universal profiles for $\Froude^*\rightarrow\infty$, each color in the image corresponds to one Froude number. (\emph{a}) at $\tau=2$, all shapes collapse. (\emph{b}) at $\tau=3$, a difference becomes visible for $\Froude^*=2.7$ and $\Froude^*=5.3$ (\emph{c}) $\tau=4$, for increasing $\Froude^*$, the shapes of the entrained oil converge to a single universal profile independent of $\Froude^*$. ($d$-$f$) shows the same profiles (only retaining shapes fulfilling the universality criterium $\tau^2 < Fr^*$) together with the boundary integral simulation results (dotted lines). The position and size of the disc is indicated by the black rectangle. The profile of the oil-water interface is not shown in the vicinity of the disc because the profile was not visible due to the vortex ring, see Fig. \ref{fig:sequence_thick_layer}.
    }
\end{figure}

Figure \ref{fig:sequence_thick_layer} shows two experiments in which we pulled down the disc from the oil-water interface at different velocities $V_D = 0.25$ and $1.0~\mathrm{m/s}$. Initially the disc was at rest and its lower surface was aligned with the oil-water interface. Then, the disc was set into motion and in a short period of time obtained a constant speed $V_D$ \footnote{With an acceleration of $42~\mathrm{m/s^2}$, it takes $0.024~\mathrm{s}$ to reach $V_D=1.0~\mathrm{m/s}$ ($\Froude^*=43$). The duration of the experiment in that case is $0.08~\mathrm{s}$. The acceleration does not significantly influence the experiment, as can be appreciated in Fig.~\ref{fig:all_profiles}}. A vortex ring appeared just above the disc, along with a smooth profile in the center which connects to the thick oil layer at the top. We first focus on this smooth profile, which has similarities with the shapes seen in \cite{Lian1989}, although that study  concentrated on the formation of the vortex ring.

In order to compare the experiments for different disc speeds, we define a dimensionless time $\tau=z(t)/R_D$, where z(t) is the depth the disc has reached at time $t$ after starting from $z=0$ at $t=0$, i.e., at equal dimensionless times the disc has reached the same vertical position below the undisturbed oil-water interface, measured in units of the disc radius $R_D$. By varying the acceleration we verified that at equal target velocities $V_D$ and dimensionless times $\tau$ the observed column shapes are independent of the precise value of the acceleration used in experiments~\footnote{We have used accelerations ranging from $21$ to $42~\mathrm{m/s^2}$ and found no discernible differences in the shape of the entrained oil column, up to the point in time that gravity becomes significant, where small differences would be introduced due to the somewhat longer time that is needed to reach a depth $z(t)$ at small accelerations.}. If the disc would be moving at a constant velocity $V_D$ all the time, i.e., if the acceleration phase would be infinitely small, then $\tau$ and $t$ would be related as $\tau=tV_D/R_D$. In the remainder of the article we will for simplicity ignore the existence of the acceleration phase and take $z(t) = V_D t$.

Comparing Fig.~\ref{fig:sequence_thick_layer}(\emph{a}i) and (\emph{a}ii), we see that at $\tau=2$ the shape of the entrained oil is very similar for $\Froude^*=2.7$ and $\Froude^*=43$, although the amount of vorticity in the vortex ring appears to be 
much larger for the higher speed. At $\tau=3$ the effect of buoyancy becomes visible for the lower Froude number, where a difference in the shape of the entrained oil between Fig.~\ref{fig:sequence_thick_layer}(\emph{b}i) and (\emph{b}ii) is appreciable. In the last frame, Fig.~\ref{fig:sequence_thick_layer}(\emph{c}i) and (\emph{c}ii) at $\tau=4$, the oil in the case of $\Froude^*=2.7$  has clearly moved back up due to buoyancy, leaving only a relatively straight cylinder of oil behind. For $\Froude^*=43$, the shape is still unaffected by gravity at this point in time.

In order to see what the effect of the density difference is on the shape that we obtain at high Froude numbers, we performed experiments with demineralized water ($\rho=998~\mathrm{kg/m^3}$) at $V_D=1~\mathrm{m/s}$ ($\Froude^*=99$) and with salt water ($\rho=1140~\mathrm{kg/m^3}$) at $V_D=1.5~\mathrm{m/s}$ ($\Froude^*=97$). Although there is a factor two in the density difference between the oil and the water phases (demineralized vs. salt water), the difference in the profiles is negligible, which leads us to conclude that the shape of the entrained oil column does not strongly depend on the relative density difference between the fluids. The use of salt water does however have an experimental advantage: The oil-water interface became less contaminated with oil and water droplets after the experiment is finished, which reduced the time that we had to wait between two experiments until the surface was smooth enough to clearly observe formation of the profile of the entrained oil. For experimental convenience, we used salt water in all experiments, except noted otherwise.

In Fig.~\ref{fig:all_profiles} we compare the profiles of the entrained oil at equal dimensionless times $\tau$ and for a wide range of Froude numbers. Every experiment in Fig.~\ref{fig:all_profiles}(\emph{a-c}) consists of several repetitions of the experiment, sometimes using two different disc accelerations ($21$ and $42~\mathrm{m/s^2}$), indicating the excellent reproducibility of the experiment and the irrelevance of the initial startup motion of the disc and the actual acceleration that is used in this startup phase.

The appearance of differences in the shapes shown in Fig.~\ref{fig:all_profiles}(\emph{a-c}) are a result of gravity that is pushing the oil phase upwards. This will only happen if the time is long enough for gravity to become more important than the inertia that is pulling the oil phase down. We can predict the moment that differences appear by comparing the inertial time scale $t_{in}\equiv R_D/V_D$ to the gravitational time scale $t_g \equiv\sqrt{R_D/g^*}$. Gravity is expected to play a role only if $t\gtrsim t_g$, which, after dividing both sides by the inertial time scale can be written as
\begin{equation}
    \tau^2\gtrsim\Froude^*,
    \label{eq:OilTimeScales}
\end{equation}
where we have used $t/t_{in}=\tau$. If we now again look at Fig.~\ref{fig:all_profiles}(\emph{a-c}), we expect according to Eq.~(\ref{eq:OilTimeScales}) to see a difference for $\Froude^*\lesssim4$ at $\tau=2$, for $\Froude^*\lesssim9$ at $\tau=3$, and for $\Froude^*\lesssim16$ at $\tau=4$. These predictions agree well with the moment that we observe differences in the experimental profiles in Fig.~\ref{fig:all_profiles}(\emph{a-c}).

We now proceed to compare our experimental findings to potential flow solutions. For this, we use the boundary integral method as described in~\cite{Oguz1993,Bergmann2009,Gekle2010d,Gekle2011a,Pozrikidis2011}. We performed boundary integral simulations of an impulsively started disc in an infinite bath of fluid, moving at a constant speed $V_D$. The disc had the same size and thickness as in the experiment. We injected tracers at the position corresponding to the initial oil-water interface, \textit{i.e.}, aligned with the bottom of the disc. The tracers were then advected with the flow field around the disc. Because we used tracers in an infinite bath of a single liquid, the motion of the tracers corresponds to the case of $Fr^*\rightarrow\infty$. To validate the numerical code, we verified our simulations by calculating the displaced volume in the case of a sphere. We found good agreement with the analytical results of Eames~\textit{et al.}~\cite{Eames1994} for $\rho_{max}/|x_0|\rightarrow\infty$, where $\rho_{max}$ is the radius of the reference plane that is taken into account in the calculation of the drift volume and $x_0$ the initial axial distance of the sphere from this plane (see Eames~\textit{et al.}~\cite{Eames1994} for details). We note that in our experiments the initial position of the disc is at the reference plane (the oil-water interface), such that $x_0\rightarrow0$, and, consequently, $\rho_{max}/|x_0|\rightarrow\infty$.

Fig.~\ref{fig:all_profiles}\emph{d-f} shows that the shapes are approximated by our simulation results, but also that there exists a significant difference close to the undisturbed surface. A closer inspection reveals that the discrepancy increases as the disc moves further down: while at $\tau=2$ ($d$) there is a reasonable agreement between the simulation and experiment, at $\tau=4$ ($f$) the difference is much larger. The same discrepancy is also illustrated in Fig.~\ref{fig:sequence_thick_layer}($a$ii-$c$ii).

\begin{figure}
    \centering
    \includegraphics{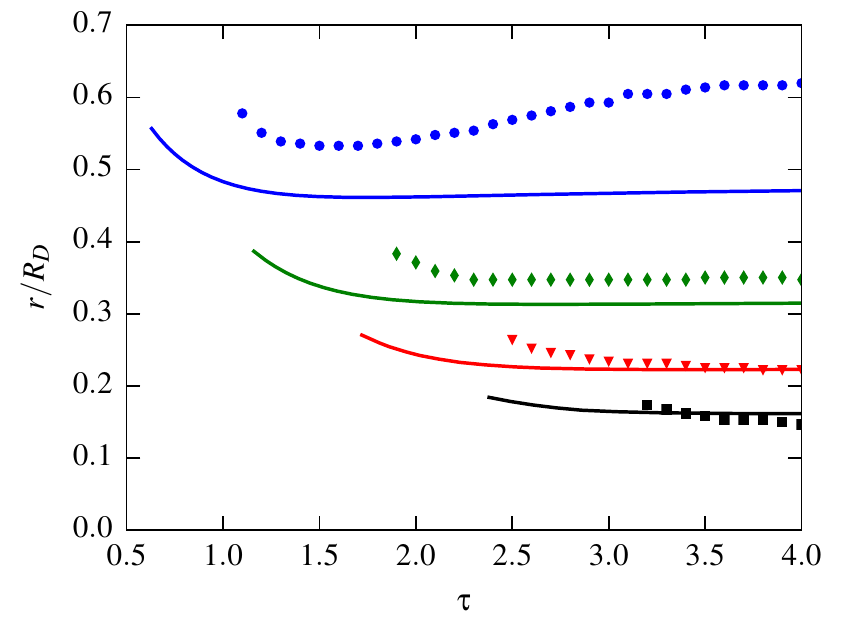}
    \caption{\label{fig:radiusTime}
    Radius of the entrained column oil as a function of dimensionless time $\tau$ at different depths $z$ for the experiment (closed symbols) and the simulation (solid lines). Blue circles: $z/R_D=-0.5$, green diamonds: $z/R_D=-1.0$, red triangles: $z/R_D=-1.5$, black squares: $z/R_D=-2.0$.}
\end{figure}

We further investigate the development of this difference between the experiments and simulation in Fig.~\ref{fig:radiusTime}, where we plot the radius of the oil/water interface at different depths $z$ below the location of the undisturbed interface at $z=0$. First, we observe that the larger the distance to the undisturbed surface is, the smaller the discrepancy becomes. The difference in radius even appears to switch sign at the deepest position plotted here (black squares), although this difference is close to the experimental uncertainty. Second, the figure shows an apparent qualitative difference between the experiments and simulations at $z/R_D=-0.5$, for which the simulation shows an initial decrease of the radius, which then seemingly asymptotes to a constant value. The experiment on the other hand shows a significant increase in radius after the initial decrease~\footnote{Actually, also in the simulations we observe such an increase, only much smaller ($<2\%$ of the minimum radius) than in the experiment}.

A qualitative explanation for both these observations originates from the formation and growth of the vortex which introduces an additional downward velocity (and velocity gradient) inside the entrained volume of oil, since oil will continuously flow into the vortex. There where the shape of the entrained oil volume $r(z,t)$ is column-like (i.e., $\partial r/\partial z$ is small), this will lead to simple stretching and thus a decreasing radius of the column~\cite{Eggers2008}. However, where there exists a large gradient in the radius, the overall downward translation may introduce an increase in radius in the lab frame. This can be seen as follows. Because the volume $\propto r^2v_z$ of the entrained oil is conserved, the interface $r(z,t)$ obeys the PDE
\begin{equation}
	\frac{\partial r^2}{\partial t} + \frac{\partial}{\partial z}(r^2 v_z)=0,
\end{equation}
where $v_z(z,t)$ ($\geq 0$) is the downward velocity in the vertical direction inside the column, which for simplicity is assumed to be independent of the radial coordinate. This can immediately be rewritten as
\begin{equation}
	\frac{\partial r}{\partial t} + v_z\frac{\partial r}{\partial z} + \frac{1}{2}r\frac{\partial v_z}{\partial z}=0.
\end{equation}
Note that for such a stretching flow the second term in the above expression is always negative, since $\partial r/\partial z < 0$ for the measured profiles, whereas the third (or stretching) term is always positive because $\partial v_z/\partial z \geq 0$. In case of a purely columnar shape, the second term is zero, and therefore $\partial r/\partial t$ is negative. In case of a funnel-shaped profile with a strong radial gradient ($\partial r/\partial z \ll 0$) and sufficient downward velocity, the magnitude of the second term may well
be larger than the stretching term, which will result in a positive $\partial r/\partial t$.

The remaining question is why we still observe universal shapes even though there is a clear influence from the vortex. The reason is that for early times ($\tau\lesssim8$) the
dimensionless size and strength of a vortex behind a impulsively started disc is independent of the disc speed, as was shown in~\cite{Yang2012}. That the same holds for our experiments can clearly be seen in Fig.~\ref{fig:sequence_thick_layer}, where the vortex for $Fr^*=2.7$ has the same size as that for $Fr^*=43$, when they are compared at the same dimensionless time.

\begin{figure}
    \centering
    \includegraphics{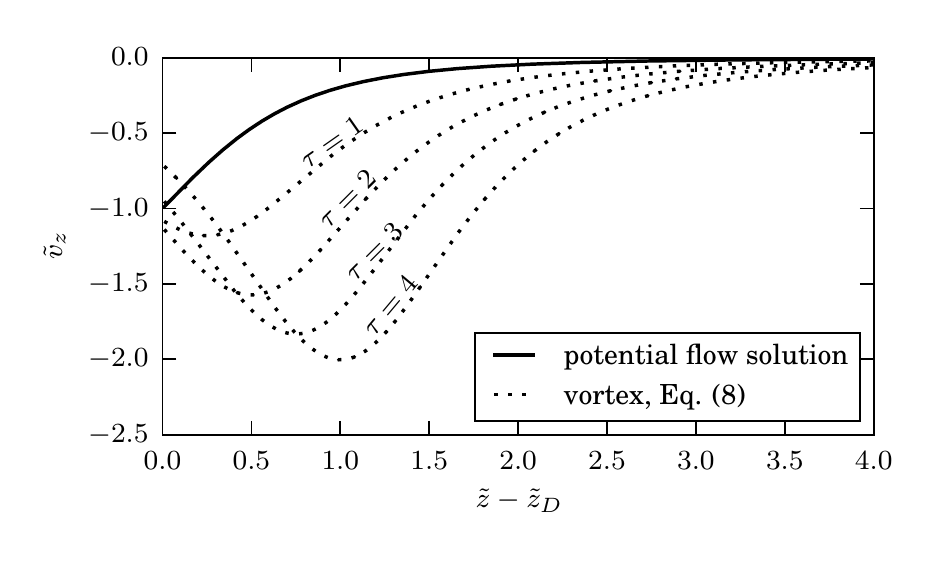}
    \caption{\label{fig:vzVortex}
    Comparison of the estimated non-dimensional velocity $\tilde v_z = v_z/V_D$ on the axis of symmetry as a function of the non-dimensional distance $\tilde z -\tilde z_D = (z-z_D)/R_D$ to the center of the disc for the potential flow solution (solid line) and the time-dependent contribution from the vortex ring (dotted lines) at four different instances in dimensionless time $\tau = tR_D/V_D$.}
\end{figure}

We now proceed to providing an estimate of the magnitude of the downward velocity $v_z$ due to the presence of the vortex. Given the circulation $\Gamma$ of a vortex line in a closed loop $C$, the velocity $\bf v$ at position $\bf r$ can be obtained from the Biot-Savart law
\begin{equation}
	{\bf v}({\bf r}) = \frac{\Gamma}{4\pi}\oint_C\frac{d{\bf l}\times({\bf l} - {\bf r})}{|{\bf l} - {\bf r}|^3},
	\label{eq:BiotSavart}
\end{equation}
where $\bf l$ is the core of the vortex line. For a ring-shaped vortex with a radius equal to the disc radius $R_D$ we can write for points $\bf r$ on the axis of symmetry that, taking into account that the velocity is purely vertical, $|{\bf l} - {\bf r}|=\sqrt{R_D^2+(z')^2}$ and ${\bf \hat{e}_z} \cdot (d{\bf l}\times({\bf l} - {\bf r})) = R_D dl$, with $\bf \hat{e}_z$ the unit vector in the $z$-direction, $dl=R_Dd\theta$, and $z'=z-z_c$, where $z_c$ is the vertical position of the core of the vortex. The $z$-component of the integral of Eq.~(\ref{eq:BiotSavart}) can now be evaluated straightforwardly as
\begin{equation}
	v_z(z') = \frac{\Gamma}{4\pi}\int_0^{2\pi}\frac{R_D^2}{(R_D^2 + z'^2)^{3/2}}d\theta
\end{equation}
and gives the vertical component of the velocity $v_z$ as
\begin{equation}
	v_z = \frac{R_D^2\Gamma}{2(R_D^2+z'^2)^{3/2}},
\end{equation}
or, in dimensionless form,
\begin{equation}
	\tilde v_z = \frac{\tilde\Gamma}{2(1+\tilde z'^2)^{3/2}},
\end{equation}
where $\tilde v_z = v_z/V_D$, $\tilde z = z/R_D$, and the dimensionless, time-dependent circulation $\tilde\Gamma=\Gamma/(R_DV_D)$ is independent of the disc speed~\cite{Yang2012}. In Fig.~\ref{fig:vzVortex} we compare the vertical velocity on the axis of symmetry resulting from our potential flow calculation to the one induced by the vortex ring at different instances of time. We have used the empirical relation from Yang~\emph{et al.}~\cite{Yang2012} for $\tilde\Gamma(\tau)$~\footnote{$\tilde\Gamma\approx-4.67(1-0.75e^{-0.416\tau} - (1-0.75)e^{-24.927\tau})$} and approximated the velocity of the vortex core to be linear (i.e., $\tilde z'=\tilde z_D + 0.25\tau$). The latter approximately matches the position of the vortex core in the experiments, where we observe that the core is about one disc radius above the disc at $\tau=4$ (see Fig.~\ref{fig:sequence_thick_layer}). Clearly, the velocities and velocity gradients are greatly enhanced by the starting vortex, and an influence on the shape of the entrained oil column is therefore expected. Most importantly note that the velocity profiles of Fig.~\ref{fig:vzVortex} are independent of the disc velocity, resulting in a mechanism through which the universality of the shapes shown in Fig.~\ref{fig:all_profiles} are preserved.

To further quantify the influence of the vortex on the drift volume, we calculate the entrained volume of oil in both the simulations and the experiment. Because a part of the interface is masked by the vortex, we do not calculate the complete displaced volume~\cite{Eames1994}, but only compare the part that is accessible in both the experiment and the simulation. In the simulations we find, at $\tau=4$, between the depths $z/R_D=-0.2$ and $z/R_D=-1.0$ (with $z=0$ at the unperturbed oil/water interface), an entrained volume of 4.24 ml, while in the experiment 7.25 ml of oil is entrained. Extending the range to $z/R_D=-2.0$ gives 5.57 ml and 8.66 ml for the simulation and experiment respectively. Clearly, the experimental volume of entrained oil is significantly larger than the volume predicted by the potential flow simulations.

\section{Conclusions and outlook}
We have performed experiments where we started a disc at an oil-water interface and pulled it down at a constant speed. We have shown that at high speeds, gravity and surface tension can be neglected, and the  entrained oil obtains a universal funnel shape, independent of the Froude number. However, a vortex ring is formed at the disc edge, which influences the shape of the entrained oil resulting in a qualitative and quantitative difference compared to the potential flow solution. Nonetheless, the universality of the funnel shape is conserved. The shape also appears insensitive to the relative density difference, at least for the density differences studied.

The effect might be investigated further for flows at lower Reynolds numbers, exploring the limit where the vortex disappears. In our current setup this regime is inaccessible due to the influence of gravity at low disc speeds. A consequence of our finding is that the displaced volume as predicted by \citep{Darwin1953} is underestimated in cases where flow separation is of importance. This introduces a non-trivial shape-effect on the entrainment in the wake of, for example, ocean life \cite{Katija2009}. The preserved universality in such cases, however, may help in simplifying and generalizing analysis in these situations.

\begin{acknowledgments}
We acknowledge financial support from FOM and the NWO-Spinoza program.
\end{acknowledgments}

\end{document}